\documentclass[a4, 10 pt, conference]{ieeeconf}  
\IEEEoverridecommandlockouts                              
\usepackage{graphicx}
\usepackage{booktabs}
\usepackage[hyphens]{url}
\newcommand{\customfootnote}[1]{%
    \begingroup
    \renewcommand{\thefootnote}{}
    \footnote{#1}%
    \addtocounter{footnote}{-1}
    \endgroup
}

\title{\LARGE \bf
AsyncMLD: Asynchronous Multi-LLM Framework \\ for Dialogue Recommendation System
}
\author{Naoki Yoshimaru$^{1}$, Motoharu Okuma$^{2}$, Takamasa Iio$^{2}$ and Kenji Hatano$^{2}$
    \thanks{*This work was partially supported by JSPS KAKENHI Grant Number 19H05691, 22H03895 and 23H03694.}
    \thanks{$^{1}$ Graduate School of Culture and Information Science, Doshisha University, 1-3 Tatara-Miyakodani Kyotanabe, Kyoto 610-0394, Japan}
    \thanks{$^{2}$ Faculty of Culture and Information Science, Doshisha University, 1-3 Tatara-Miyakodani
    Kyotanabe, Kyoto 610-0394, Japan}
}

\begin{document}
\maketitle
\thispagestyle{empty}
\pagestyle{empty}

\begin{abstract}
We have reached a practical and realistic phase in human-support dialogue agents by developing a large language model (LLM). 
However, when requiring expert knowledge or anticipating the utterance content using the massive size of the dialogue database, we still need help with the utterance content's effectiveness and the efficiency of its output speed, even if using LLM.
Therefore, we propose a framework that uses LLM asynchronously in the part of the system that returns an appropriate response and in the part that understands the user's intention and searches the database.
In particular, noting that it takes time for the robot to speak, threading related to database searches is performed while the robot is speaking.
\end{abstract}
\section{Introduction}\label{sec:Intro}
The demand for dialogue agents that can support humans is increasing daily; in particular, the emergence of the Large Language Model (LLM) in recent years has dramatically advanced research, and services such as ChatGPT and BingAI have made dialogue systems more 
accessible to various people \cite{lee-2021-improving-end}.
In Task-Oriented Dialogue systems (TOD), where specific tasks or specialized and complex conversations are conducted in the dialogue, the challenge in using LLM is to process the conversation while performing multiple tasks. 
This challenge includes, for example, extracting information for actual automatic ticket sales or searching a large database and modifying the conversation based on the results. 
To perform these processes, it is necessary to dynamically change multiple components while keeping track of the dialogue situation, and it is not easy to do so using only prompts from an LLM.

Therefore, we propose a framework in which the appropriate dialogue response and the data retrieval part of the recommendation target are divided, and the dialogue proceeds asynchronously using two LLMs. 
This system makes it possible to combine advanced reasoning with natural dialogue.
Specifically, we focus on the fact that it takes time for the robot to speak through speech synthesis, and while one thread process is speaking, another thread process understands the speech and searches the database. 
These processes are asynchronously connected to each other and are synchronized before the next conversational turn takes place, so the interaction is smooth in terms of response speed.
Thus, our framework can be used for various dialogue systems with databases, but this paper describes a case in a travel agency.

\section{Proposed Framework}\label{sec:Proposed}
We propose a framework that can effectively promote dialogues by asynchronous coordination of modules when a massive amount of data is recommended. 
This is an asynchronous operation that separates (A) the system that appropriately interacts with the user and (B) the system that understands the intent of speech and performs searches in the database.

\begin{figure}[t]
    \centering
    \includegraphics[width=\linewidth]{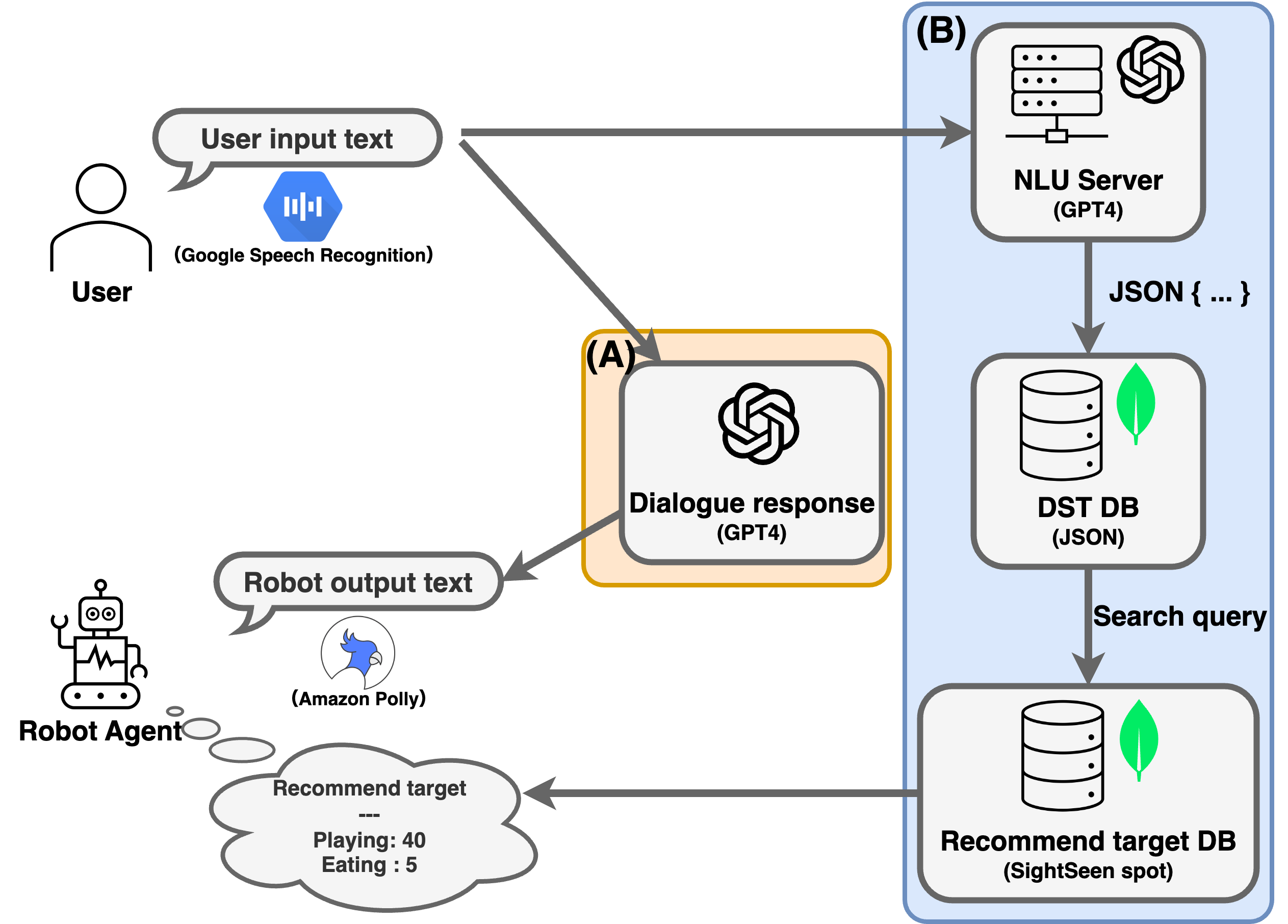}
    \caption{\textbf{Architecture of AsyncMLD:} It consists of four main components, and (A) and (B) are processed asynchronously to advance the conversation. The brackets under each component describe the technology used in this case. }
    \label{fig:proposed_flow}
    \vspace{-6.5mm}
\end{figure}

\subsection{Asynchronous processing}
In our system, part(A) and part(B) in Fig. \ref{fig:proposed_flow} are processed asynchronously. 
First, the user's voice is converted to text, then the text is passed to (A) and (B) simultaneously.
In (A), the dialogue response component moves, and the LLM responds appropriately. 
In (B), the NLU server receives the user's text and converts it to JSON format. The data is stored in the DST DB, and the current dialogue state is updated. After that, a search is performed in the RT DB based on the DST information, and the next conversation is decided based on the recommendation status.

\subsection{System component}
The specific architecture is shown in Fig. \ref{fig:proposed_flow} and consists of the following four components:

\begin{LaTeXdescription}
    \item[Dialogue response component:] This part mainly interacts with the user. 
    After converting speech data into text, the data is input to the language model for output. 
    The GPT4 \cite{openai2023gpt4} language model is used in this system, and the prompts are set to dialogue sentences that encourage recommendations.
    \item[NLU component:] After converting the user's speech input into text, the text is input to the language model to understand the user's intentions. It designed the prompt to output slots matching the DB data. For example, if there are metadata (major categories: [Playing, Seeing, Eating, ...]) assigned to a tourist attraction.
    \item[DST DB component:] Dialogue intentions obtained from users are stored in the dialogue State Tracker Database (DST DB). It accumulates the data where NLU is performed, and elements are added without duplication.
    \item[RT DB component:] The results obtained from NLU are used in a search for the Recommend Target DB (RT DB). In this case, the database stores information on tourist attractions with meta information.
\end{LaTeXdescription}

\subsection{Implementation}
We implemented the system described above in the following environment. 
Docker is used for efficient coordination between modules, especially in asynchronous processing, and NLU and MongoDB containers are always running to facilitate dialogue.
\begin{itemize}
    \item Dialogue progression, NLU Server: Python
    \item DST DB, RT DB: MongoDB
    \item Speech recognition: Google Speech Recognition 
    \item Voice generation: Amazon Polly
\end{itemize}

\section{Sightseeing Spot Recommendation}\label{sec:Desc_DRC2023}
Our system was tested at the Dialogue Robot Competition 2023\cite{DRC2023} (DRC2023).
In this competition, the system was operated by a customer who actually visited travel agencies in Nagoya and Fukuoka, recommended two tourist attractions, and presented a plan to visit them. 
The demonstration experiment was conducted using the following interaction procedure by the regulations.

\begin{enumerate}
    \item Start Dialog: Speak welcome statement
    \item Sightseeing Spot Recommendation using AsyncMLD: In a natural dialogue, our system selected four sightseeing spots from more than 100 spots in Kyoto.
    \item Choice of two Sightseeing Spot: Explain to the user the details of the four tourist attractions and ask him to choose two of them.
    \item Create a route between two Sightseeing Spot: Create transportation routes using NAVITIME API \footnote{NAVITIME API, \url{https://api-sdk.navitime.co.jp/api/}, (accessed December 14, 2023.)}
    \item Question and answer session: Provide LLM prompts with information on which Sightseeing Spot to visit and the user's response
    \item End of Dialog: Speak a statement announcing the end of the dialog
\end{enumerate}

\section{Result on DRC2023} \label{sec:DRC2023_res}
In DRC2023, this system was evaluated by a questionnaire with 21 subjects, although the number of people who experienced the system varied depending on the implementation dates of each team.
After the users had experienced the system, they were asked to answer nine evaluation questions to measure their satisfaction with the interaction and two questions to measure their satisfaction with the plan.
The results are shown in Table \ref{tab:result_tab}.

\begin{table}[t]
    \centering
    \caption{Competition Result}
    \vspace{-2mm}
    \begin{tabular}{crr}
        \toprule
        Team & Dialogue experience & Travel plan \\
        \midrule
        Baseline & 38.68 \; (7st)  & 0.6667 \; (5st)\\
        Ours & \textbf{44.62 \; (3rd)} & 0.7700 \; (8st) \\
        \bottomrule
    \end{tabular}
    \vspace{-5mm}
    \label{tab:result_tab}
\end{table}

The level of satisfaction with the dialogue exceeded the baseline. In particular, the item that evaluates whether the user can refer to the information obtained from the robot when choosing a sightseeing spot performed the best, scoring 5.57 points. 
The results of the operation showed that the robot was able to respond quickly in conversation with the customer, despite the multiple components in motion.
The fact that it scored 0.74 higher than Baseline for the question "Did you interact with the robot naturally?" suggests that it can provide more appropriate responses, including response speed, than methods using a single LLM.
On the other hand, the evaluation of the travel plan was low, especially in the area of subjective judgment on whether it was realistic or not. 
This may be attributed to the fact that the route information was given one-sidedly in step 4 of section \ref{sec:Proposed}; although it is factual information obtained from the API, it is not in a format that can be understood by the user.
Since plans to visit several sightseeing spots tend to be complex, we will generate responses in such a way that users can understand them smoothly and without confusion.

\section{Conclusion}\label{sec:Concle}
We proposed a dialogue system framework that draws out objects to be recommended from the dialogue. 
One of the issues to be addressed is the generation of prompts on the NLU component. 
In the current system, metadata is obtained from the data to be recommended and input into the prompts. 
Still, the number of characters may exceed the number of characters allowed by the prompts as the size of the data increases. 
Therefore, in the future, we will dynamically change the prompt generation to reduce the burden on the language model.
Since plans to visit several sightseeing spots tend to be complex, we will generate responses in such a way that users can understand them smoothly and without confusion.
In addition, uses open-source LLMs to measure scalability and facilitate realistic discussions of time and cost costs.

\customfootnote{
AUTHOR CONTRIBUTION: Conceptualization, Implementation, and Management, N.Y; 
Prompt Engineering, N.Y, and M.O; 
Writing Paper, N.Y; 
Supervision T.I and K.H. 
}

\bibliographystyle{IEEEtran}

\vspace{-10mm}
\begin{thebibliography}{1}
\bibitem{lee-2021-improving-end}
Y. Lee, “Improving end-to-end task-oriented dialog system with a simple auxiliary task,” in Findings of the Association for Computational Linguistics: EMNLP 2021. Punta Cana, Dominican Republic: Association for Computational Linguistics, 2021, pp. 1296–1303.
\bibitem{openai2023gpt4}
OpenAI, “Gpt-4 technical report,” 2023.
\bibitem{DRC2023}
M. Takashi, H. Ryuichiro, S. Kurima, F. Tomo, N. Hiromitsu, and N. Takayuki, “Overview of dialogue robot competition 2023,” in Proceedings of the Dialogue Robot Competition 2023, 2023.
\end{thebibliography}

\end{document}